\documentclass[superscriptaddress, 
reprint,
amsmath,amssymb,aps,]{revtex4-1}

\usepackage{graphicx}
\usepackage{dcolumn}
\usepackage{bm}

\begin{document}
\large
\title{\Large  Arresting the collapse of a catenary arch}

\author{Jemal Guven}
\email{\emph{E-mail address:} jemal@nucleares.unam.mx}
\affiliation{Instituto de Ciencias Nucleares, Universidad
Nacional Aut\'onoma de M\'exico,  
04510 Ciudad de M\'exico, MEXICO}
\author{ Gregorio Manrique}
\email{\emph{E-mail address:} gregorio.manrique@correo.nucleares.unam.mx}
\affiliation{Instituto de Ciencias Nucleares, Universidad
Nacional
Aut\'onoma de M\'exico,  
04510 Ciudad de M\'exico, MEXICO}

\begin{abstract}
It is well known that viable architectural structures can be identified by locating the critical points of the gravitational potential energy congruent with some fixed surface metric.
This is because, if the walls are thin, the lowest energy modes of deformation are strain-free, and thus described by surface isometries.
If it is to stand, however, an arch had better possess some minimum rigidity. The bending energy consistent with this construction protocol, we will show,  can only depend on curvature deviations away from the reference equilibrium form. 
The question of stability, like the determination of equilibrium, turns on the geometry. 
We show how to construct the self-adjoint operator controlling the response to deformations consistent with isometry.  As illustration, we reassess the stability of a simple catenary arch in terms of the behavior of the ground state of this operator. The energy of this state increases monotonically
with the bending rigidity and  it is possible to identify the critical rigidity above which the arch is rendered stable. 
While this dependence may be monotonic, it exhibits a number of subcritical kinks indicating significant qualitative changes in the ground state associated with eigenvalue crossovers among  
the unstable modes in the spectrum; the number of such competing modes increasing rapidly as the rigidity is lowered.  The initial collapse of a subcritical arch is controlled by  the ground state; on the critical threshold,  there are two unstable modes of equal energy, one raising the arch at its center, the other lowering it. The latter dominates as the instability grows. The qualitative behavior of the ground state changes as the rigidity is lowered---its nodal pattern as well as its parity undergoing abrupt changes as the intervals between crossovers converge---complicating the  prediction of the dominant initial mode of collapse.

\end{abstract}


\maketitle

\section{Introduction}  

The first people to build with adobe, brick or stone would have been quick to appreciate that these materials tend to withstand compression considerably better than they do tension \cite{Hey95}.
Unreinforced masonry will resist substantial compressive forces due to gravity before it crumbles; it will, however, develop cracks when subjected to even modest tension. But,  before it fails, it will behave--to a surprisingly good approximation---as an inextensible medium: it neither stretches nor shrinks. 

Contemporary building materials, we know, are not limited by tension like masonry, and they will buckle rather than crumble. Nevertheless, the most relevant modes of deformation consistent with the integrity of any thin-walled free-standing structure,  
whether it be a cathedral vault or the canopy of an airport terminal, are the same as those that constrain the design of masonry.  For, if the walls are thin compared to their radius of curvature, the modes costing the least energy are those preserving distances along the surface of the wall.  In mathematical terms, this means that they leave the metric on this surface fixed: they are isometries.
The subject of isometry was discussed with (unusual) relish in volume five of Spivak \cite{Spivak}.  The curious reader will find complementary treatments in  references \cite{Pogorelov,Harnach,Audoly} and, perhaps most comprehensively, in Steven Verpoort's thesis \cite{Verpoort}.
Not only are the relevant degrees of freedom geometrical, so also are the constraints connecting them.
Despite the difference in scale, in this sense, the walls of any building---viewed in the large---are no different from a folded thin sheet, 
be it graphene or 
a modest A4 sheet of paper.  The latter has come under increasing  scrutiny in recent years not only for its technological importance but also as a legitimate problem---not to be underestimated---falling within the scope of soft matter  \cite{TomWitten}; for a sampling of the rapidly expanding literature see, for example,
\cite{Maha,Sharon,Starostin,Gemmer}.  Technically, even the parametrization of an isometrically folded thin sheet, never mind the physics, poses its own challenges (as emphasised forcefully in a recent paper \cite{Fried}).  In retrospect, the neglect of isometry by physicists or engineers until the closing years of the twentieth century  will require some explanation. 

We will argue that, 
despite the differences in the underlying physics and the disparity in scale involved, the geometrical framework used to describe thin sheets can be tweaked to shed interesting light on the behavior of contempory building materials and, whenever the thin wall idealization is appropriate, the behavior of masonry itself.  Paradoxical as it may seem, there is a limit in which it is not unrealistic to think of masonry as a soft material. 

The recognition 
that a catenary somehow provides the optimal structure for spanning an opening  predates Robert Hooke \cite{TaqKasra}. 
He was, however,  the first to articulate why in the language of mechanics; 
indeed,  the significance of his discovery impressed him sufficiently that he squandered valuable time composing an anagram in Latin to record it \cite{Hooke,Wiki,Osserman}.  Deciphered and translated, the well-known anagram declaims {\it ``As hangs the chain, so but inverted stands the arch''}. 
A hanging
rope of fixed length,
suspended between two fixed points,
will minimize its potential energy
when it traces out a catenary; this part is an elementary exercise in the calculus of variations accommodating the local constraint on its length. Inverted, it will form an arch. In the process, the
tension in the chain gets converted to compression and because the lines of thrust are tangent to the catenary they are steered safely into the foundations or lateral supports \cite{Roman}.
Curiously, in contrast to the law reluctant schoolchildren associate with Hooke, there is not a strain in sight. 

More generally, if the gravitational potential energy is stationary with respect to deformations
preserving the metric, the stress will be tangential and the surface viable.
Remarkably, this is true irrespective of its specific material properties. The problem is completely geometrical.  The equations leading to this conclusion can be derived by applying the calculus of variations using Lagrange multipliers to impose isometry as a constraint. We will show how to do this following the method developed by one of the authors (with Martin M\"uller) in another context \cite{GM08}.  

Despite the simplicity of Hooke's protocol, it remains the basis for identifying rather more complex optimal geometries, with custom-designed software enlisted to implement it in its full two-dimensional glory. The extraordinary scope of the approach is illustrated  in references \cite {VHWP12}  and \cite{ShellStructures}. Its aesthetic potential was, of course,  anticipated by 
Antoni Gaud\'\i{},  and more recently by the late Zaha Hadid
\cite{Gaudi,Zaha}.  
One could quibble over the occasionally awkward aesthetics---or question the impulses driving the construction of such cavernous structures---but such questions fall well outside the scope of this paper. There is, however, a crucial aspect of the protocol that is not addressed in the optimization process: an assessment of the stability of these structures in the geometrical spirit of the protocol \cite{Displacements}.This turns out not to be straightforward;  and geometry plays an even more significant role than it does in the design itself. 


The stress, of course, is not constant within an arch.  In Hooke's protocol, however, it depends only on the geometrical aspect ratio and its mass; it does not depend on any other material properties. Notably, 
resistence to bending is ignored---with impunity---in the design protocol;  so it cannot contribute to the stress distributed within the equilibrium structure. 
This may be very well in a hanging rope but, in an arch, it cannot be the full story. For 
the protocol also ignores the fact that the gravitational potential energy  is  a maximum in a catenary arch. 
As such, it would collapse immediately, like a falling rope
whether or not the one-third rule of thumb---confining the lines of thrust---is satisfied \cite{Hey77,LeviSalvadori02}.  
Unlike a pencil, standing on its lead, the number of unstable  modes of the arch consistent with isometry is infinite (or certainly large); the modes with the shortest wavelength turn out to be the most unstable gravitationally. 
While the cost to bend the arch may play no role in determining the equilibrium shape, or even contribute to the stress, it clearly does play a very important role in counteracting this instability.  

But this begs the question: what is the appropriate bending energy? 
The energy consistent with Hooke's
protocol cannot be the familiar symmetric energy quadratic in
curvature: for, if it were, the equilibrium would itself possess bending energy which would in turn shift the equilibrium, 
introducing additional bending stresses within the structure. 
Hooke's protocol would be rendered as good as useless. 
The appropriate bending energy can only penalize deviations away from the equilibrium {\it reference} state, at lowest order it will be quadratic in the deviation. There is no bending energy in the equilibrium state. This is not only true in  
the traditional construction of a masonry arch which involves the use of centring,  removed when the cement has cured, but also in the assembly of pre-formed units in contemporary construction where the accommodation of 
tension is not an obstacle.  Indeed, this understanding is implicit in a promotional video produced by Norman Foster and Partners, where the principle informing the design of an airport terminal is illustrated using a chain with resin sprayed onto the hanging chain to lend it the rigidity it needs to stand  upon inversion \cite{Fostervideo,wikipaediaMexico}. The bending energy possesses an inhomogeneous spontaneous curvature determined by the equilibrium we spray. 
 
Bending energies with this property may vanish in the equilibrium state but  will contribute at second and higher order in deformations about it. It is also manifestly positive.  If sufficiently positive, the arch will stand.

For simplicity, we will suppose that a single parameter $\mu$ characterizes the bending rigidity of the medium. 
This simple energy is completely geometrical.  Thus the question of stability, like the Hookean construction itself,  is a geometrical question. 
Extensions to accommodate anisotropies or inhomogeneities are all straightforward in principle, but obscure the essential  geometrical nature of the problem.

To examine stability under small deformations it is necessary to expand the total energy---gravitational potential plus bending energy---constrained by 
isometry, to quadratic order about the equilibrium structure. It is not obvious that the inclusion of the constraint at this order
is going to be tractable analytically. By happy accident, we find it turns out to be.  The key result of this paper is the 
identification of the self-adjoint linear operator controlling small deformations and thus the stability of the structure. Technically there is a snag: 
this operator is sixth-order in derivatives---a consequence of the constraint; an order that one is not accostumed to chance upon in the description of physical systems. 

To illustrate how our framework may be applied, we will examine a symmetric catenary arch and assess its stability from this point of view. This is as simple as it gets yet not quite as simple as we would have hoped. The compensation is that it turns out to be considerably more interesting than
expected. 
 
If the parameter $\mu$ is positive the spectrum of the operator is bounded from below.  There is also a critical rigidity $\mu_c$ above which  the lowest eigenvalue turns positive and the arch is rendered stable with respect to small deformations;  below $\mu_c$ the arch will be unstable; its  
dominant mode of collapse described by the eigenfunction of this operator which  corresponds to this lowest eigenvalue.  To this point the behavior accords with intuition.  The lowest eigenvalue also grows montonically with $\mu$. However,  as it grows, it displays kinks, signalling interesting behavior where these occur.

It is useful to 
think of the self-adjoint operator as  a Hamiltonian describing a quantum 
mechanical particle moving in one-dimension---albeit an unusual Hamiltonian sixth order in momentum---and to identify the eigenfunction which corresponds to the lowest eigenvalue  as the ground state of this Hamiltonian.  

When the arch is just subcritical, the ground state is symmetric and exhibits two nodes. This single state represents two physically distinct  unstable modes: one raising the arch at its center, the other
lowering it.  This degeneracy cannot be resolved at quadratic order and it is necessary to proceed to the next order in deformations to identify which of the two is the more unstable. It turns out to be the mode lowering the arch at its center and this mode will dominate as the instability grows, consistent with observations. The hinge-points where failure occurs are located at the three anti-nodal positions along the arch.  This interpretation of this feature appears to be new. We predict how the positions of these hinges depend on the aspect ratio as well as the rigidity.  
 
The spatial behavior of the ground state depends somewhat sensitively on how far below criticality the rigidity lies,  indicating a  qualitative change in the instability  as the rigidity is reduced. Below a second critical value, located at the last kink,  
the unstable symmetric state with two nodes is replaced by an antisymmetric state with an additional node as the ground state. The ground state is no longer the state with the lowest number of nodes and it is no longer symmetric.  
As the rigidity is lowered even further, the number of nodes displayed by the ground state generally increases as one passes through the
succession of discrete values of $\mu$, where the kinks are located.  The number of unstable states also increases.  What is more,  
the number of nodes exhibited by these states does not correlate simply with their energy which, unlike the ground state energy, does not generally increase monotonically with $\mu$. 
Predicting the mode of collapse is  complicated accordingly. The collapse of a simple catenary arch, as simple an architectural structure as one can conceive, turns out to be an unexpectedly complex process. 

It is possible that the reader  may not be interested in this specific problem.
However,  the issues addressed here and their treatment are increasingly relevant in soft matter and the recently dubbed field of extreme mechanics where isometries play an important role in understanding
both the morphology and the physical behavior of thin sheets, whether macroscopic or nanoscaled.  The framework we introduce can be adapted  to accommodate different bending energies as well as external forces.


\section{Gravitational potential and bending energy}
If the structure is thin compared to its radius of curvature, we
can represent it as a parametrized surface in three-dimensional
space, $(u^1,u^2)\mapsto \mathbf{X}(u^1,u^2)$. Here $\mathbf{X}=(X^1,X^2,X^3)$ is a triplet of functions, providing a Cartesian representation of the surface. 
The gravitational potential energy  is given by
\begin{equation}
\label{potl}
H_0[\mathbf{X}] = \rho  \int dA\, h 
\,,
\end{equation}
where 
$h=\mathbf{X}\cdot \mathbf{k}$ is the local height above a given horizontal plane, and $\mathbf{k}$ is the normal to this plane. The measure of surface area $dA$ is invariant under isometry.
We will suppose for simplicity that the mass density $\rho$ is
uniform and absorb the acceleration due to gravity
into its definition.
The Hookean equilibria are critical points of $H_0$ within the 
class of geometries with fixed boundaries and a given induced
metric.
The latter is given by 
\begin{equation}
\label{gdef}
g_{ab}=\mathbf{e}_a\cdot \mathbf{e}_b\,,
\end{equation}
where
$\mathbf{e}_a=\partial_a\mathbf{X}$, $a,b=1,2$ are the two surface 
tangent vectors adapted to the parametrization. We will examine the corresponding equilibria in 
Section \ref{Hooke EL}.
\\\\
{\it Equilibrium biased bending energy:}
The bending energy of a surface is quadratic in its curvature;
if it is symmetric and isotropic, and the reference geometry is flat,
it is given by (see, for
example, \cite{Landau})
\begin{equation}
\label{HB0}
H_{B0}= \frac{1}{2} \mu  \int dA \, (C_1+C_2)^2 +
\bar\mu \int dA\, C_1 C_2\,.
\end{equation}
Here $C_1$ and $C_2$ are the principal curvatures,  namely, the two real
eigenvalues of the extrinsic curvature tensor, 
a measure of how fast the surface
normal $\mathbf{n}$ rotates into one direction as one moves
it along another,
\begin{equation}
\label{Kdef}
K_{ab}= \mathbf{e}_a\cdot \partial_b \mathbf{n}\,.
\end{equation}
The Gaussian curvature $C_1 C_2$ is an isometry
invariant.
Modulo the two material 
parameters, $\mu$ and $\bar \mu$,
this energy depends only on the geometry. If sufficiently thin 
(of thickness $t$), 
we also know that the surface will bend before it stretches \cite{Ray90}: this is because 
the bending energy scales as $t^3$, whereas stretching energy increases linearly with $t$. We will comment on potential limitations of this approximation in the conclusions.

In the Hookean construction, the appropriate bending energy,
$H_B$,  cannot play any role in determining the equilibrium nor in
determining the distribution of stress.  As such, both $H_B$ and
its first variation, $\delta H_B$, must
vanish in equilibrium. Neither condition is satisfied by $H_{B0}$. 
Nor is the reference geometry flat. 
The simplest energy consistent with these criteria is given by
\begin{equation}H_B = \frac{1}{2}\,\mu\, \int dA\, \sum_I (C_I-
{\cal C}_I(u))^2 \,,
\label{HB}
\end{equation}
where 
${\cal C}_1[u]$ and ${\cal C}_2[u]$ are the 
curvatures of the undeformed reference geometry,  the equilibrium determined by  the minimization of the gravitational potential energy consistent with the constaints. Because we are only interested in isometries there is no need to introduce a term analogous to the second
term, proportional to $\bar\mu$,  appearing in Eq.(\ref{HB0}).

The two curvatures ${\cal C}_1[u]$ and ${\cal C}_2[u]$ appear to play the role of anisotropic, position dependent. spontaneous curvatures. However, it would be wrong to think of them as independent parameters in the model, determined as they are  by the equilibrium we are deforming. Constant
isotropic spontaneous curvatures are familiar in 
membrane biophysics where they are iintroduced to explain the global morphology of cellular membranes \cite{Helfrich,Seifert}; they can also arise due to the differences in the area of the two sides of a membrane bilayer \cite{SvetinaZeks89}, or indeed a difference in the areas of the two bilayers within a laminar tetralayer enclosing a lumen, as conjectured in \cite{GHV14}. Energies involving constant anisotropic spontaneous curvatures  occur in two-dimensional liquid crystals;  curved nematics on a fluid membrane substrate  may imprint their curvature on that substrate \cite{FrankKardar, Walani}.  Spatially varying anisotropic target curvatures are also relevant in the modelling of the controlled lateral swelling of thin elastic sheets \cite{Dias}.
What distinguishes the energy (\ref{HB}) from these is the fact that  
${\cal C}_1[u]$ and ${\cal C}_2[u]$ are themselves undetermined until the Hookean optimization is completed. This is a very important feature of the model \cite{noflanalog}. 
While $H_B$ vanishes in equilibrium, it does contribute at second and higher order in departures away
from equilibrium. It is positive and thus tends to
stabilize the geometry.  There are two questions we wish to answer: how large must $\mu$ be to achieve this? What is the nature of the instability if $\mu$ falls short?  As we have hinted, and as we will show, the answer to the latter question presents a number of surprises. 

\section{Hooke's shape equations}
\label{Hooke EL}
The critical points of the energy can be identified by using the
method of Lagrange multipliers to impose the isometry
constraint.  To solve the Hookean problem, one introduces the
constrained functional (the metric is defined in terms of $\mathbf{X}$ by Eq.(\ref{gdef}) and 
$g_{ab}^{(0)}(u^1,u^2)$ is a fixed 
metric of our choosing),
\begin{equation}
H_0{}_C[\mathbf{X},\sigma^{ab}] = \rho \int dA\, h - \frac{1}{2}
\int dA \, \sigma^ {ab} (g_{ab}- g_{ab}^{(0)})\,.
\end{equation}
A familiar analogue is the treatment of vector fields with a
fixed---usually vanishing---divergence.
To determine the Euler-Lagrange equations, we examine how the
energy responds to a surface deformation $\mathbf{X}(u^1,u^2)\to
\mathbf{X}(u^1,u^2)+ \delta\mathbf{X}(u^1,u^2)$. The presence of the local multipliers $\sigma^{ab}(u^1,u^2)$
frees us to treat the three Cartesian deformations, $\delta
\mathbf{X}$, independently, unconstrained by isometry \cite{GM08}. 

Qualitatively, normal and tangential deformations are very
different. As such, it is useful to decompose the deformation explicitly along the respective directions:
\begin{equation}
\label{delXdecomp}
\delta\mathbf{X}= \Psi^a \,\mathbf{e}_a+ \Phi\,\mathbf{n}\,.
\end{equation}
The
normal and tangent Euler-Lagrange equations are
identified using methods developed elsewhere (we use  
\cite{CGS,GM08}):
\begin{subequations}
\label{EL}
\begin{eqnarray}
(\sigma^{ab} - \rho h\,g^{ab})\, K_{ab} = \rho \,
(\mathbf{n}\cdot \mathbf{k})\,;\\
\nabla_a\, \sigma^{ab}=0\,.
\end{eqnarray}
\end{subequations}
Here $\nabla_a$ is the covariant derivative compatible with
$g_{ab}$; the normal equation involves the extrinsic curvature
$K_{ab}$ defined in Eq. (\ref{Kdef}). In general, Eqs.(\ref{EL}) would 
underdetermine the geometry if $g_{ab}$ were not fixed.  The element of freedom  in the design process is captured
mathematically by the choice of metric. 
Eqs.(\ref{EL}) are, themselves, well-known. They appear in 
references \cite {VHWP12} and \cite{ShellStructures}. While neither derivation involves the calculus of variations this is, however, the obvious approach if one is to proceed and assess the stability of the equilibrium geomerty. 

Whereas the stress associated with the constraint
is conserved, the
total stress $T^{ab}= \sigma^{ab}- \rho h \, g^{ab} $, adding
the contribution from gravity, is not: $\nabla_a T^{ab}= -\rho
(\mathbf{e}^b\cdot \mathbf{k})$. It is the total stress, however, that needs
 to be  consistent with the  properties of the medium;
in masonry, it must be compressive everywhere (or
approximately so); 
equivalently, $T^{ab}$ must be positive definite.
\\\\
{\it Symmetric catenary vault:}  
As a prelude to the assessment of its stability, we re-examine briefly---in this approach---a cylindrical vault described in an adapted  Cartesian parametrization, $(X,Y)$, by $h= h(X)$. This is not, of course,  the  most straightforward approach if one is only interested in equilibrium.  Indeed, an elementary, if incomplete, derivation  using force balance is provided at the beginning of Gray's graphic introduction to differential geometry  \cite{Gray}.

Eq.(\ref{EL}b) implies that
$\sigma_{XX}=\sigma_0$, a constant; whereas $\sigma_{YY}=0$.
Using the identification of $K$ with the Frenet curvature along the curve described by $h$,
$\kappa= -\ddot h/ (1+ \dot h^2 )^{3/2} $, where dot denotes $
d/dX$, Eq.(\ref{EL}a) integrates to yield a quadrature
$\dot {\tilde h}^2 - C^2 \tilde h^2 =-1 $, where we define $\tilde h := h -
\sigma_0/\rho$. This posseses the familiar solution:
$\tilde h = C^{-1} \cosh C X$. The two unknown constants $C$ and $\sigma_0$
are determined by the boundary conditions.
For a cross-section
of maximum height $H$, and width $2L$, one determines $\sigma_0$ 
implicitly as a function of the aspect ratio: 
\begin{equation}
\sigma_0 / \rho = ( \sigma_0 /\rho - H) \cosh ( L/
[\sigma_0/\rho - H])\,.
\end{equation}
This relationship is represented in
Fig. 1. It is now straightforward to read off the compressive stress at any position along the arch.
\begin{center}
\includegraphics[scale=0.40]{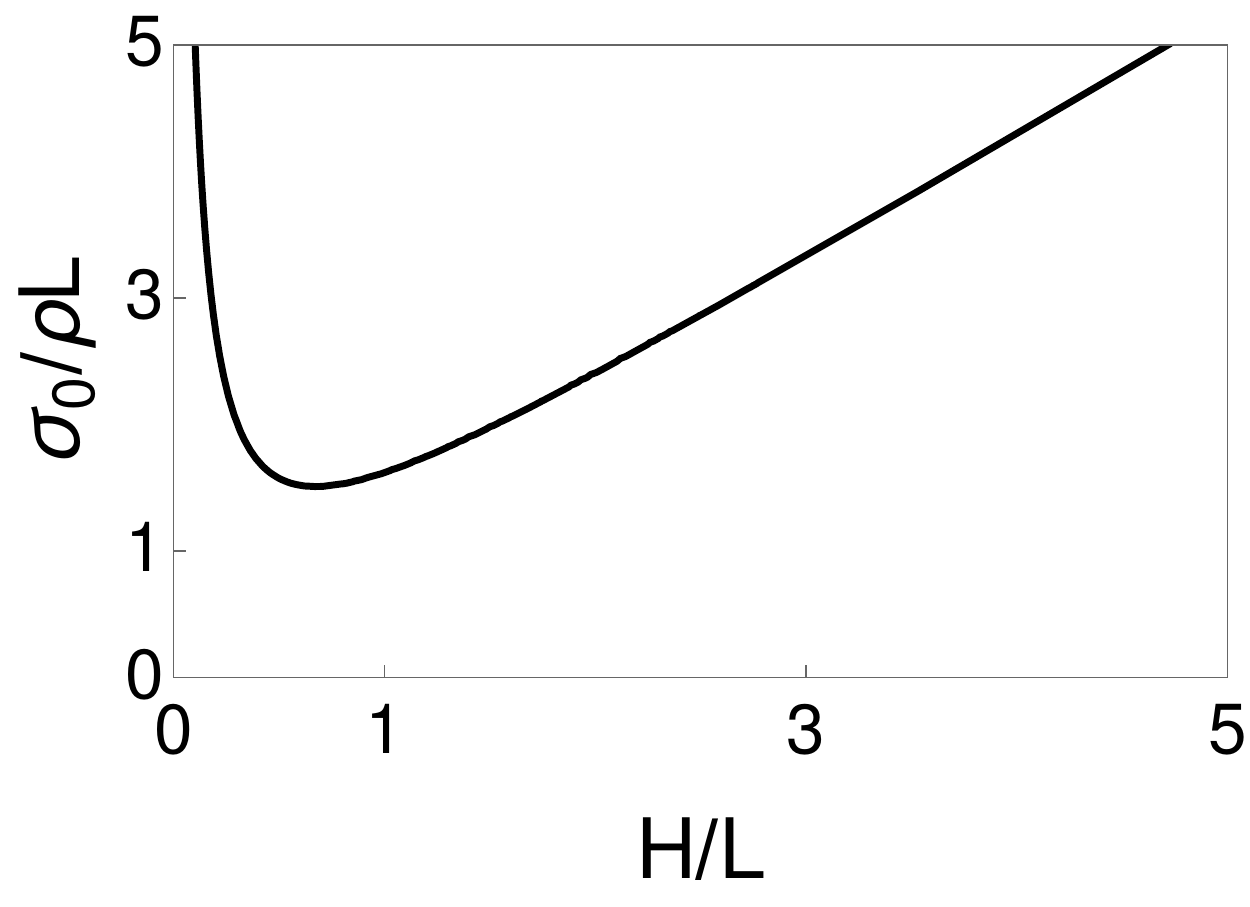}
\end{center}
\noindent{\small {\bf FIG. 1}: The stress associated with the isometry constraint as a function of the aspect ratio in a catenary arch. $\sigma_0$ is positive; in addition, $\sigma_0/\rho>  H$  so that the total stress $T_0=  \sigma_0 - \rho  h  > 0$ is compressive everywhere.}

\section{ Second variations}

We now turn to the question of stability of the equilibrium structure.
This involves expanding the total energy, the sum of  the potential energy 
(\ref{potl}) and bending energy (\ref{HB}) to second order in deformations  about this structure. Again these deformations need to be consistent with isometry.  We will do this in general, for fixed boundaries.
The correct boundary conditions on free boundaries is a vexed issue that will be addressed elsewhere.
\\\\
 {\it (1) Potential energy}
One can show that the second variation of the constrained
potential energy is given by (we abuse notation by writing $H_0$ instead of $H_{0C}$) 
$\delta^2 H_0 = 
- \int dA \, T^ {ab} \,\nabla_a \delta \mathbf {X} \cdot
\nabla_b \delta \mathbf {X} $.
Because the total stress $T^{ab}$ is positive in compression, it follows that
$\delta^2 H_0 < 0$. The deformation $\delta\mathbf{X}$ appearing here 
is constrained by isometry, or equivalently, at this order: $\mathbf{e}_a\cdot \nabla_b\delta\mathbf{X} +\mathbf{e}_b\cdot \nabla_a \delta\mathbf{X}=0$.
This constraint can be
recast at linear order in terms of the projections of $\delta
\mathbf{X}$ (\ref{delXdecomp}) as
\begin{equation}\label{isol}
\nabla_a\Psi_b + \nabla_b \Psi_a + 2 K_{ab} \,\Phi =0\,.
\end{equation}
This allows 
$\delta^2 H_0$ to be rewritten  
\begin{equation}\label{d2H0}
\delta^2 H_0 =  -
\int dA\,  
T^ {ab} [ A_a A_b + \frac{1}{4} g_{ab} (\nabla \times
\Psi)^2]\,,\end{equation}
where we introduce the surface vector field 
\begin{equation}
\label{Adef}
A^a := \nabla^a
\Phi - K^{ab} \Psi_b\,,
\end{equation}
and define $\nabla\times \Psi=
\epsilon^{ab}\nabla_a \Psi_b$ using the antisymmetric
Levi-Civita tensor $\epsilon^{ab}$. Each of the two terms appearing in 
Eq.(\ref{d2H0}) is manifestly negative. 
\\\\ {\it (2) Bending energy}
If $\delta \mathbf{X}$ is an isometry, the second variation of
the equilibrium biased bending energy (\ref{HB}) is given by 
\begin{equation}
\delta^2 H_B= \mu\,  \int dA\,\sum_I  (\delta C_I)^2\,,
\end{equation}
where $\delta C_I$ is the deformation induced in the curvature $C_I$ under the isometry.
We now use the trivial algebraic identity 
$2 C_1 C_2 = (C_1+C_2)^2 - C_1^2 - C_2^2$ to express the Gaussian curvature as a difference.  
As a consequence, under isometry, $\sum_I (\delta C_I^2)  = (\sum_I \delta C_I)^2$. But $\sum_I C_I = g^{ab} K_{ab}$. Thus
\begin{equation}
\delta^2 H_B= \mu\,  \int dA\, (\delta K)^2\,.
\end{equation}
Remarkably, it is only necessary to determine the deformed trace to evaluate the second variation of the bending energy. This can be shown to be given, for an
isometry, by 
$\delta  K 
= - \nabla_a (\nabla^a \Phi - K^{ab} \Psi_b) = - \nabla_a A^a$, where $A^a$ defined in Eq.(\ref{Adef}) appears once again. This follows from the general expression 
$\delta K=  - ( \nabla^2  + K^{ab} K_{ab} )\Phi + \Psi^a\partial_a K$ (see, for example, \cite{CGS}), and the isometry constraint, Eq. (\ref{isol}).

One can now add the two
contributions to identify the total second variation,
$\delta^2 H_{\sf total} =
\delta^2 H_B  + \delta^2 H_0$. 
This expression is general.  Indeed, we do not need to be interested specifically in gravity. Our analysis continues to be valid with an obvious modification if gravity is replaced by a more general external force derivable from a potential $V$, with the replacement $h$, linear in $\mathbf{X}$, by $V(\mathbf{X})$. 

We are specifically
interested in the critical value of the rigidity $\mu$
above which $\delta^2 H_{\sf total} > 0$. We will show how
to determine this value for a symmetric catenary arch.

\section{Stability of the catenary arch}

We first examine deformations preserving the cylindrical symmetry. The second variation now simplifies to give 
\begin{equation}
\delta^2 H_{\sf total}=
\int ds\, \left[\mu \,A'^2 
  -  (\sigma_0 - \rho h) A^2 \right] \,,
 \end{equation}
 where $A= \Phi' -\kappa\Psi$ is now  a scalar,  and again 
  $\kappa$ is the Frenet curvature of the catenary.
$\Psi$ is the projection of 
$\delta\mathbf{X}$ along the unit tangent, $\Phi$ again is the normal deformation, and prime indicates a
derivative with respect to arc-length.
The two fields $\Phi$ and $\Psi$ are not independent,  
constrained as they are by the cylindrical reduction of 
Eq.(\ref{isol}):  $\Psi'+ \kappa \Phi=0$. 
Note that the second term appearing in Eq.(\ref{d2H0}) does not contribute. It does, however,  contribute if the deformation is not cylindrical, but curls will also contribute to the bending energy 
so, a non-vanishing curl, does not necessarily imply that a higher rigidity is required 
to establish stability.

The natural way to proceed might appear to be to eliminate
$\Psi$ in favor of $\Phi$, as done in \cite{GMV}. After all, intuitively,  one tends to think of the tangential deformations as tagging their normal counterparts, not the other way around. The fixed boundary 
conditions, however,  obstruct this course. 
Counterintuitive as
it may seem, they do permit $\Phi$ to be eliminated in
favor of $\Psi'$ and this can be done globally because $\kappa>0$. 
We now can express the scalar $A$ in terms of $\Psi$: $A={\cal F}\, \Psi$, where the differential operator
${\cal F}$ is given by 
\begin{equation}
 {\cal F} := -\partial_s \frac{1}{\kappa} \partial_s - \kappa \,.
 \label{Fdef}
 \end{equation}
Because the boundary is fixed, $\Psi=0$, 
$\Psi'=0$, and $\Psi''=0$ there. One can now integrate by parts
to rewrite
\begin{equation}
\delta^2 H_{\sf total}=
\int ds\,  \Psi \mathcal{H}_\mu \Psi\,,
\end{equation}
where $\mathcal{H}_\mu$ is self-adjoint on the appropriate Hilbert
space of functions. This operator is
the sum of a
positive definite sixth-order operator $\mathcal{H}_1$
associated with bending energy and a
negative definite fourth-order $\mathcal{H}_0$ associated with the
gravitational potential:
\begin{equation}
\mathcal{H}_\mu= \mu \mathcal{H}_1+\mathcal{H}_0\,;
\end{equation}
where 
\begin{subequations}
\label{L1L0}
\begin{eqnarray}
&&\mathcal{H}_1= -  {\cal F} \, \partial_s^2 \, {\cal F}\,;\\ 
&&\mathcal{H}_0 = -  {\cal F} (\sigma_0 - \rho h)  \, {\cal F}\,.
\end{eqnarray}
\end{subequations}
Each of the two operators is a manifestly self-adjoint operator sandwich formed by the operator ${\cal F}$, itself self-adjoint,  defined by Eq.(\ref{Fdef}). 

To borrow the language of quantum mechanics, the second variation $\delta^2 H_{\sf total}$ can be thought of as 
the {\it expectation value} of the operator 
$\mathcal{H}_\mu$ (the quantum mechanical {\it Hamiltonian}) in the (normalized) state $\Psi$.  In this spirit, we adopt the Dirac shorthand  \cite{dirac}
\begin{equation}
\label{del2HL}
\delta^2 H_{\sf total} [\mathbf{X}] =
\langle\Psi|\mathcal{H}_\mu|\Psi\rangle\,.
\end{equation}
\\\\
{\it Identifying the spectrum of $\mathcal{H}_\mu$:}
The operator $\mathcal{H}_\mu$ possesses a discrete
spectrum: $\mathcal{H}_\mu |\lambda_{\mu,n}^\pm
\rangle = \lambda_{\mu,n}^\pm | \lambda_{\mu,n}^\pm\rangle$,
$n=1,2,3,\dots$; because $\mathcal{H}_\mu$ is self-adjoint
its eigenvalues $\lambda_{\mu,n}^\pm $
are real and the eigenstates (or modes) $|\lambda_{\mu,n}^\pm \rangle$ corresponding to distinct eigenvalues are orthogonal. We also know that each eigenstate possesses a definite parity, indicated $+$($-$) if even (odd). This is 
because  $\mathcal{H}_\mu$ is even under reflection in the origin, sending $s\to -s$, or equivalently,  the operator $\Pi$  representing the reflection commutes with $\mathcal{H}_\mu$; as a consequence
$\Pi |\lambda_{\mu,n}^\pm \rangle = \pm |\lambda_{\mu,n}^\pm \rangle$.
Notice that the isometry constraint $\Psi'+ \kappa \Phi=0$
implies that $\Psi$ is odd whenever $\Phi$ is even. 

We order the $|\lambda_{\mu,n}^\pm \rangle$ 
by the number of nodes, $n$.
Thus $|\lambda_{\mu,1}^-\rangle$ represents an odd tangential mode with one node (its normal counterpart will be even with two nodes);
$|\lambda_{\mu,1}^+\rangle$ represents an even tangential mode with two nodes, and so on.  Even and odd parity states are orthogonal: $\langle \lambda_{\mu,n}^+|
\lambda_{\mu,m}^- \rangle=0$ for any $n$ and $m$.
In general.  this ordering differs  from the ordering by energy ($\lambda_{\mu,n}^\pm$), but when $\mu$ is very large (very small) the two orderings do correspond (anti-correspond).

One does not expect to find a simple analytical expression for 
the eigenstates or even the spectrum of $\mathcal{H}_\mu$. It is possible to perform back-of-the-envelope approximations when the number of nodes is large and we will do this.  In general, however, Rayleigh-Schr\"odinger perturbation theory is least useful where the physics is most interesting and one needs to develop a numerical approximation.  To do this, we 
expand the tangential state $|\Psi\rangle$ with respect to an appropriate complete orthonormal basis $\{|\Psi_n^\pm\rangle\,,n=1,2,3,\dots\}$  that is consistent with the boundary conditions as well as possessing a definite parity \cite{basis}:
\begin{equation}
|\Psi\rangle = \sum_{n=1,\pm}^\infty C_n^\pm \, |\Psi_n^\pm\rangle\,.
\end{equation}
These states are not themselves eigenstates of $\mathcal{H}_\mu$.
If we now expand the right hand side of Eq.(\ref{del2HL})  with respect to the $\{|\Psi_n^\pm\rangle\}$, 
we find 
\begin{equation}
\langle\Psi|\mathcal{H}_\mu | \Psi\rangle 
=
\sum_{n,m,\pm}\, C_n ^\pm C_m^\pm B_{nm}^\pm\,,
\end{equation} 
where
\begin{equation}
B_{nm}^\pm= \langle\Psi_m^\pm| \mathcal{H}_\mu |\Psi_n^\pm\rangle \,.\end{equation}
The matrix elements connecting even and odd states vanish because of the parity of $\mathcal{H}_\mu$ and the definite parity of the basis elements: $\langle\Psi_m^\pm| \mathcal{H}_\mu |\Psi_n^\mp\rangle =0$. 
Our task  is reduced to the diagonalization of two square matrices, $B_{mn}^+$ and $B_{mn}^-$. 
This identifies the eigenvalues $\lambda_{\mu, n}^\pm$ as well as the corresponding eigenstates.
\\\\
{\it The ground state energy:}
For each $\mu>0$, the operator $\mathcal{H}_\mu$ possesses a lowest eigenvalue $E_\mu$; the corresponding eigenstate is the analog of  the quantum mechanical ground state,  which we label, $|0_\mu\rangle$:
$\mathcal{H}_\mu |0_\mu \rangle = E_\mu |0_\mu \rangle$. 
The second
variation (\ref{del2HL}) is bounded from below by $E_\mu$:
$ E_\mu \le \langle\Psi|\mathcal{H}_\mu | \Psi\rangle$ for all normalized $|\Psi\rangle$. 

We are interested in the functional dependence of $E_\mu$ on $\mu$. It is straightforward to show that $E_\mu$ increases monotonically with $\mu$: if $\mu_1>\mu_2$, then
\begin{equation}
\label{mono}
\langle\Psi|\mathcal{H}_{\mu_1}|\Psi\rangle
-\langle\Psi|\mathcal{H}_{\mu_2}|\Psi\rangle
= (\mu_1-\mu_2) \langle\Psi|\mathcal{H}_1|\Psi\rangle >0\,.
\end{equation}
 We are specifically interested 
in identifying the critical value $\mu_c$,  above which $E_\mu$ turns positive, signalling the transition from instability to stability.  Once it turns positive
the inequality (\ref{mono}) ensures that it remain positive.
We now apply the numerical approximation sketched earlier to identify $E_\mu$.
Significantly,  it is not a linear function of $\mu$.  
As we will see, it is not even a smooth function of $\mu$. 
Above $\mu_c$,  however, the value of $E_\mu$ is controlled by $\mathcal{H}_1$.  This implies that, asymptotically, $E_\mu$ does depend linearly on $\mu$, consistent with our expectations.

The numerically determined $E_\mu$ is plotted as a function of $\mu$ for a representative aspect ratio in
Fig. 2. 
\begin{center}
\includegraphics[scale=0.4]{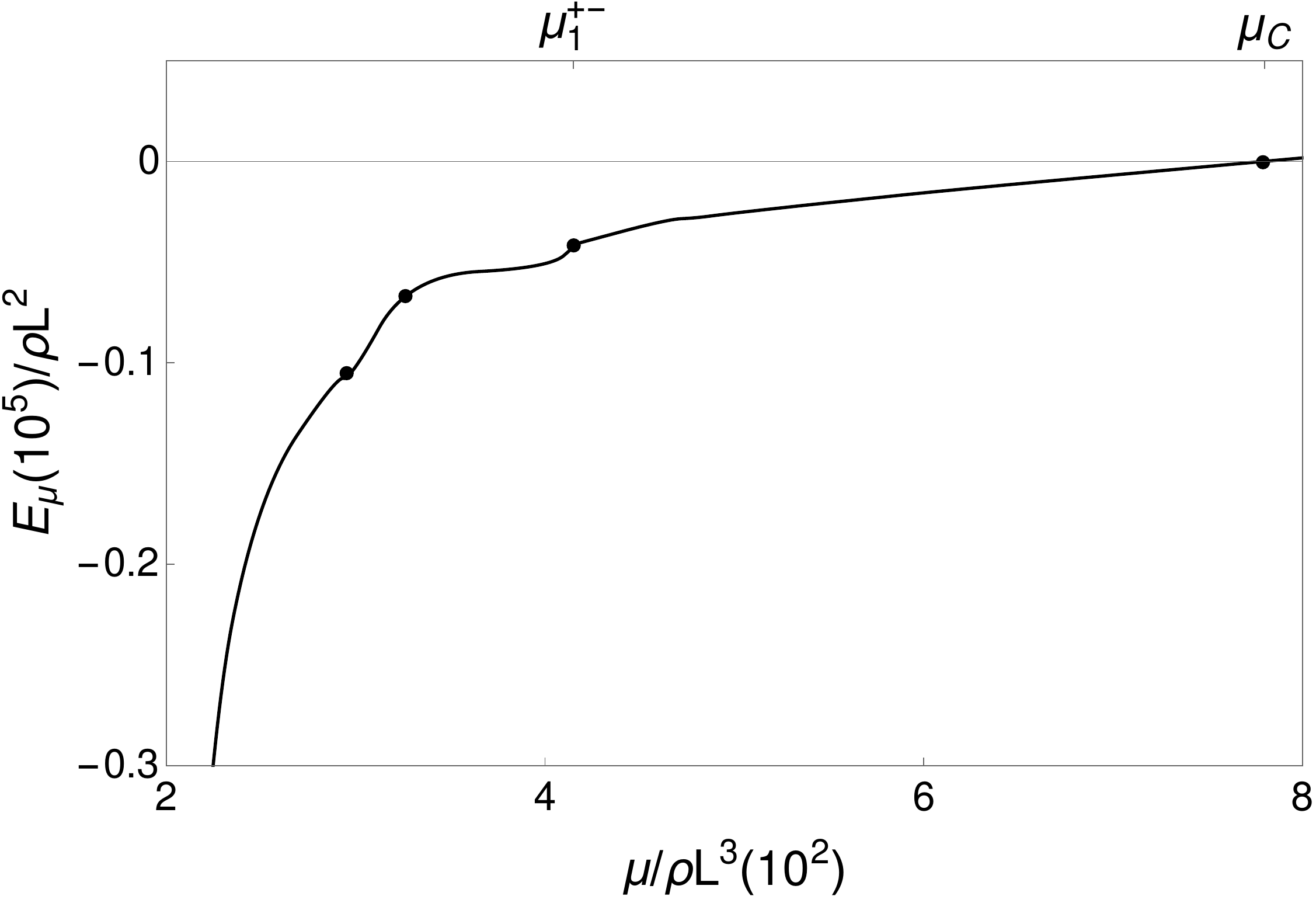}
\end{center}
\noindent{\small {\bf FIG. 2:} Ground state energy $E_\mu$ as a function
of  $\mu$ for $H/L=1/5$.  It increases monotonically with $\mu$. vanishing at the dimensionless critical value $\mu_c=7.9$ above which the arch is stable with respect to small deformations. This will be seen to occur at the value of $\mu$ ($\mu_1^-$) where the (non-monotonic) eigenvalue $\lambda_{\mu,1}^-$ changes sign for the last time.
An infinite sequence of subcritical kinks is observed in the behavior of $E_\mu$. The last three of these are indicated explicitly in the figure. The final kink occurs when $\mu/\rho L^3 (10^2)=\mu_{1}^{+-} =4.2$, where $\lambda_{\mu,1}^-$ and $\lambda_{\mu,1}^+$ crossover for the (last) time.
The origin of these kinks will be discussed in the text.} 
\vskip1pc
\noindent {\it Critical Rigidity and the aspect ratio:}  The dependence of $\mu_c$ on the aspect ratio, $H/L$ is presented in
Fig. 3. Its behavior correlates qualitatively with that of the constraining stress. We have compared these predictions with the results of 
a simple experiment involving  a thin plastic sheet suspended between two fixed walls a variable distance apart, $2L$. The initial catenary sheet collapses under inversion.  A resin lending rigidity to the sheet is now applied uniformly and the weight (a measure of $\mu$) recorded. This is repeated a number of times for each $L$ until such point that the sheet does not collapse when inverted. The results agree qualitatively with our numerical prediction. 

\begin{center}
\includegraphics[scale=0.45]{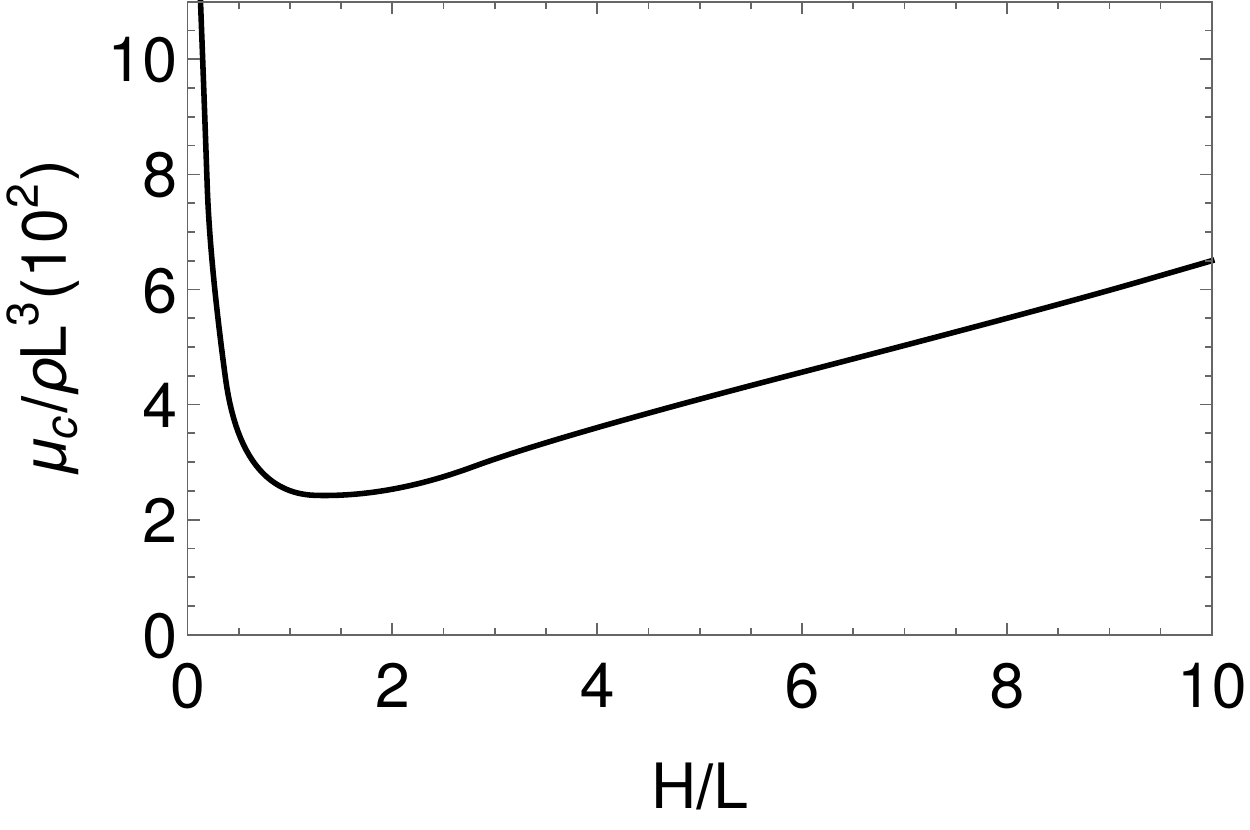}
\end{center}
\noindent{\small {\bf FIG. 3:} Critical Rigidity as a function
of the aspect ratio.}
\\\\
In a subcritical arch, the eigenstate $|\lambda_{\mu,n}^\pm \rangle$ that is assuming the role of ground state, $|0_\mu\rangle$ does not 
possess a fixed number of nodes as $\mu$ is varied, not is its parity fixed.  Because this state controls the instability of the catenary, the nature of this instability itself depends qualitatively on the value of $\mu$.  As we will see, the kinks appearing in Fig. 2 indicate the locations where these transitions occur.  To understand this behavior, 
it is necessary to study in some detail the dependence of the unstable eigenstates of $\mathcal{H}_\mu$ on $\mu$. 

As $\mu$ is decreased below $\mu_c$, the number of negative eigenvalues ---corresponding to unstable modes---typically increases.  
Let us first look at  small values of $\mu$.
In the limit of a fictitious catenary rope with no bending energy, the ground state energy in the continuum limit will be unbounded from below, corresponding to a state with an infinite number of nodes. In practice, however, there will be a short distance cutoff, so that the corresponding state is the one with the maximum number of nodes consistent with this cutoff ($n_{\rm max}$). 
\\\\
{\it Short wavelength modes:}
Suppose that the wavelength is small, or equivalently $n\gg \kappa L$, and that $H/L$ is small so that $h/L$ can be treated as slowly varying and the curvature $\kappa$ is approximately constant.  In this regime,  $\mathcal{H}_0 \approx - C_0 \partial_s^4$, where $C_0$ is a constant. As bending energy is dialled in, it will 
penalize the high curvatures in these short wavelength modes.  In the same approximation, $\mathcal{H}_1 \approx
- \mu C_1\partial_s^6$, where $C_1$ is another constant. 
The operators $\mathcal{H}_0$ and $\mathcal{H}_1$ effectively commute in this regime.  
The eigenstates of $\mathcal{H}_\mu$ can be expanded in trigonometric functions, 
with corresponding eigenvalues,   $ \lambda_{\mu\,n} \approx
C_1 n^4 ( \mu \, n^2 - C_0/C_1)$, where a  fixed geometrical factor has been absorbed  into the definition of the  constants. Note that, while the
trigonometric functions themselves are not consistent with the boundary conditions, the error can be ignored when $n$ is large.  In this regime, even and odd modes become energetically degenerate.  The eigenvalues increase linearly with $\mu$ in this regime. This is not true in general: not only do they not increase linearly, they do not even behave monotonically, the monotonic behavior of the ground state notwithstanding. 

If $\mu=0$,  $\lambda_{\mu\,n} \approx - C_0 n^4$. One sees that, were it not for the cutoff, there would be an infinite number of unstable modes consistent with isometry.

If $\mu$ is small, the lowest eigenvalue is bounded from below. If  $\mu< \mu_{\rm cut}$, where $\mu_{\rm cut}=2/3 (C_0/C_1) n_{\rm max}^{-2}$, it occurs 
when $n$ coincides with the cutoff, $n_{\rm max}$; if $\mu$ is small but exceeds $\mu_{\rm cut}$, this minimum occurs at a value $n_0 \approx \sqrt{2/3 (C_0/C_1) \mu^{-1}}$, decreasing monotonically with $\mu$. The corresponding ground state energy grows with $\mu$:  $E_\mu\approx -4/27 (C_0^3/C_1^2) \mu^{-2}$, consistent with the behavior observed in Fig. 2 for $\mu/\mu_c\ll 1$.  Notice that, unlike the low-lying eigenvalues,  $E_\mu$ does not increase linearly with $\mu$ in this regime. 

As $\mu$ is increased further and $n_0$ falls through 
single digits, this approximation becomes increasingly unreliable. One needs to account for the 
non-commutativity of the operators $\mathcal{H}_0$ and $\mathcal{H}_1$.  
We are particularly interested in the nodal behavior and the parity  of the ground state 
as $\mu$ approaches criticality. This regime does not lend itself to the simple qualitative treatment 
presented above for small values of $\mu$.

Intuitively, 
one would expect the last eigenvalue to turn positive and remain positive to be  $\lambda_{\mu,1}^-$ corresponding to the  eigenstate
$|\lambda_{\mu,1}^-\rangle$ with a single node.
Despite the unexpected erratic subcritical behavior of this state, described below, 
this expectation turns out to be justified.  
\\\\
{\it Nodal behavior of short wavelength 
modes:}  
To identify the state representing the ground state in the neighborhood of the critical rigidity, we need to understand how long wavelength modes behave in this neighborhood.  Before we do this, it is instructive to examine the behavior of short wavelength modes  in a little more detail. 
 
\begin{center}
\includegraphics[scale=0.45]{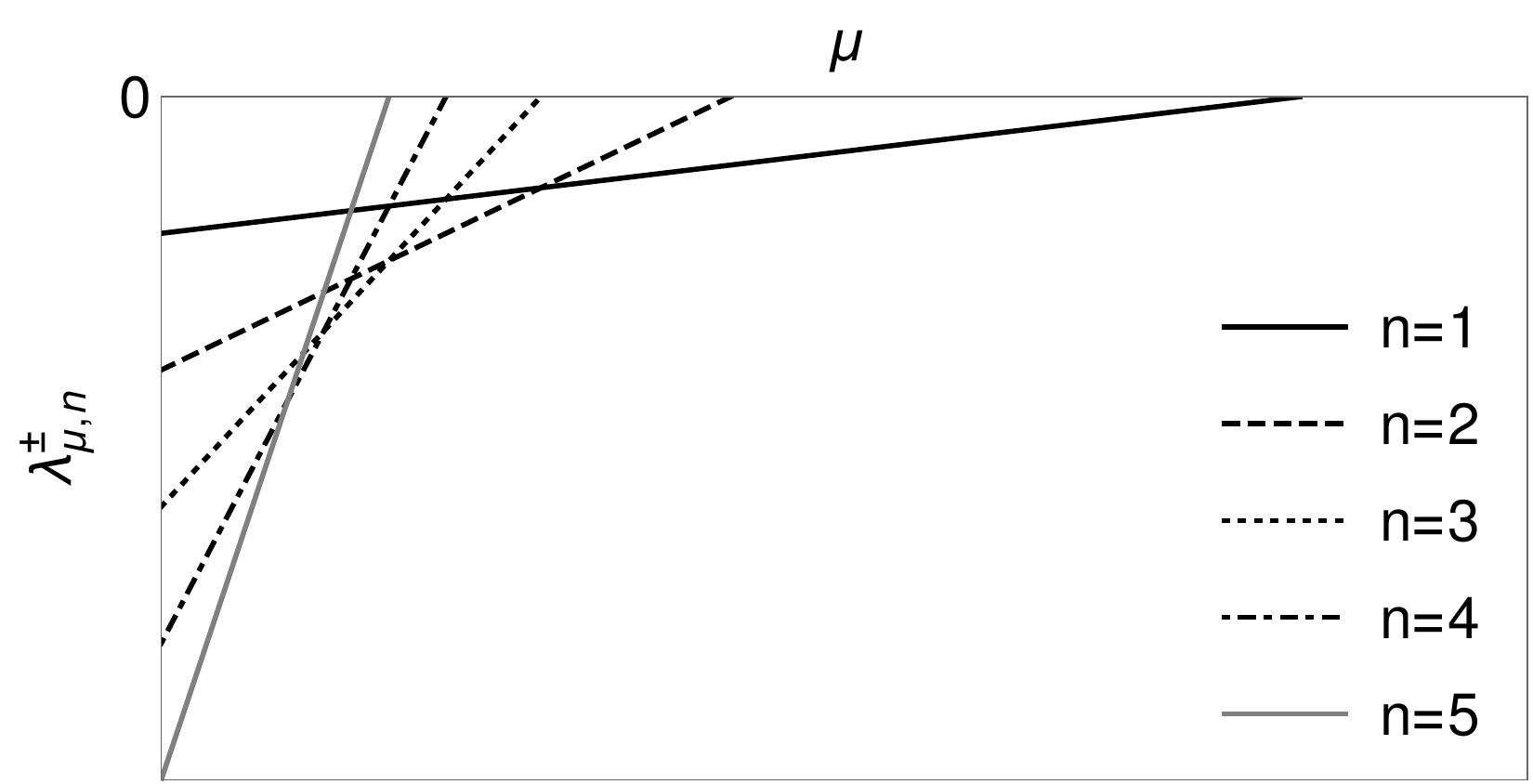}
\end{center}
{\small {\bf FIG. 4}:  $n(\mu n-1)$ vs. $\mu$,
$n=1,2,3,4,5$. This function captures the qualitative behavior of the spectrum of $\mathcal{H}_\mu$ 
when $n$ is large.
For ease of viewing, the powers of $n$ appearing here 
are not those appearing in the text.}
\\\\ 
\noindent 
Setting constants equal to one, the eigenvalues at short wavelength  
behave as 
\begin{equation}
\lambda_{\mu,n}  \sim n^4 ( \mu \, n^2 - 1)\,.
\label{lambdan}
\end{equation}
One observes the following (represented schematically in Fig.
4.):\\\\
(i) The number of unstable modes decreases with $\mu$. The
larger $n$, the more unstable the mode gravitationally; but the
high curvature in such modes also costs more bending energy so
the sooner $\lambda_{\mu,n}$ turns positive. This occurs when
$\mu_n \sim 1/n^2$.
\\
(ii) Eigenvalue crossovers (or degeneracies) occur at discrete values of $\mu$. Using Eq.(\ref{lambdan}) one finds  that, for each pair $m$ and $n$, $\lambda_{\mu,n}=\lambda_{nm}
=\lambda_{\mu,m}$  when
$\mu=\mu_{nm} = (n^4- m^4)/ (n^6- m^6)$. These crossovers occur before $\mu=\mu_n$ or $\mu_m$. Thus the corresponding
degeneracy always occurs while the eigenvalue is still negative. 
\\
(iii) The functional dependence of the ground state energy $E_\mu$ on $\mu$ is monotonic, given by the discrete envelope bounding the spectrum from below in the manner captured in Fig. 4.  Ignoring for the moment that the short wavelength expression for the eigenvalues is grossly inaccurate for small integers, we note that when $\mu>\mu_{12}$, the ground state 
consistent with the spectrum described by Eq.(\ref{lambdan}) 
is given by the state with $n=1$; when $\mu_{23} < \mu < \mu_{12}$ this state is replaced by the state with $n=2$. As $\mu$ is reduced further, the ground state is provided by
modes with increasing $n$ within ever narrowing intervals of $\mu$. 
At each transition, where the ground state is degenerate, $E_\mu$ suffers a kink. 
These are the short wavelength counterparts of the sub-critical kinks appearing in the dependence of $E_\mu$ on $\mu$ illustrated in Fig. 2.  The final kinks display  somewhat more complex behavior which we will describe more closely in the next subsection.  
\\\\
{\it  Non-Monotonicity:}
In longer
wavelength modes, the non-commutivity of the two
operators contributing to $\mathcal{H}_\mu$ cannot be ignored.
Because $\mathcal{H}_1$ is positive and $\mathcal{H}_0$ negative, the
eigenvalues of their sum, $\mathcal{H}_\mu$,
do not necessarily increase monotonically with $\mu$, never mind
linearly. This does not contradict the inequality Eq. (\ref{mono}) because  $\mathcal{H}_1$ and $\mathcal{H}_\mu$ do not possess a simultaneous of eigenstates. 
Multiple crossovers also occur. This is the origin of the striking counterintuitive non-monotonic behavior 
in the sub-critical regime displayed by $\lambda_{\mu,1}^-$ and
$\lambda_{\mu,1}^+$ in Fig. 5.
In this figure, we notice that when $\mu$ is very small, the two eigenvalues are negative,  and behave monotonically as perturbation theory would predict. In this ``unphysical'' regime they always represent the highest energy eigenvalues, not the low energy eigenvalues we are interested in.
(As we described earlier, the low energy states in this regime are  represented by short-wavelength modes.) As $\mu$  is increased further, these eigenstates undergo 
a number of large amplitude
oscillations (these are more asymmetric in the case of $\lambda_{\mu,1}^+$),  briefly
turning positive during these oscillations.  These oscillations can be understood in terms of the sensitive dependence of the energy on $\mu$ 
in longer wavelength modes associated with the non-commutativity of the two competing operators. 
This behavior itself is not unusual: it is exhibited in two-dimensional matrix models with appropriate operators. 

As $\mu$ approaches its critical value the eigenvalues again behave monotonically, signalling the domination of bending energy in these states. These excursions, remote from the critical rigidity, would appear to imply a
counterintuitive dependence of the stability on the rigidity.
Fortunately, while  $\lambda_{1}^-$ is undertaking subcritical positive excursions, the corresponding state $|\lambda_{1}^-\rangle$ is not the ground state, a role played by some other state, not necessarily $|\lambda_{1}^+\rangle$.
What this state is will depend on the aspect ratio and will require a detailed numerical examination of the subcritical spectrum beyond the scope of this work.
\begin{center}
\includegraphics[scale=0.40]{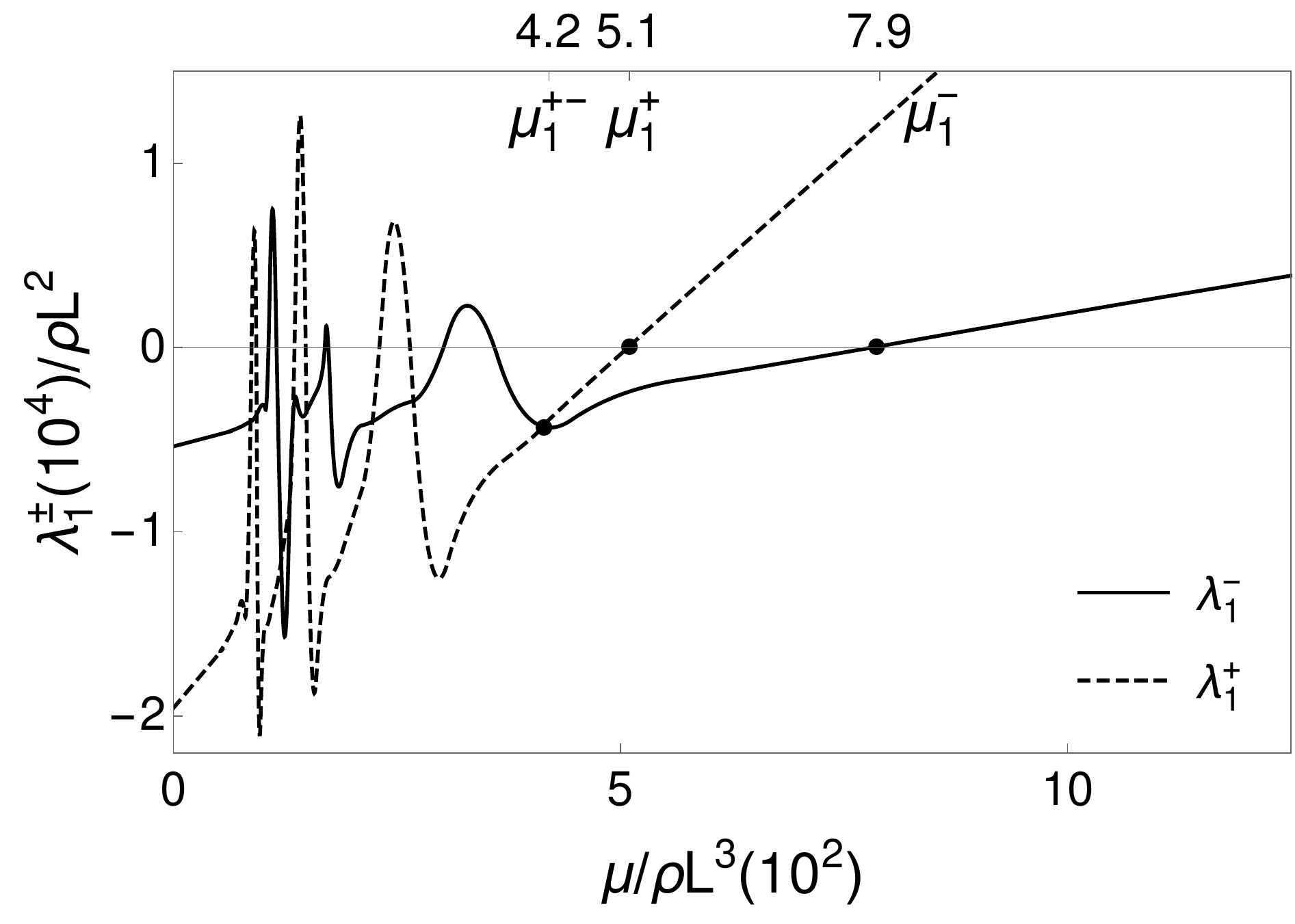}
\end{center}
{\small {\bf FIG. 5:} $\lambda_{\mu,1}^\pm$; $H/L=1/5$. If
$\mu/\rho L<<1$ then $\lambda_{\mu,1}^\pm$ both grow linearly with 
$\mu$, as perturbation theory predicts (in this regime, neither is the ground state);
if $\mu/\rho L>>1$, on the other hand, $\lambda_{\mu,1}^\pm$ are
again linear. The eigenvalues behave non-monotonically and
exhibit multiple crossovers in the subcritical regime.
The last crossover occurs at $\mu_{1}^{+-}$, $\lambda_{\mu,1}^\pm$
turn positive and remain positive at $\mu_1^\pm$. 
$\mu_1^->\mu_1^+$ and is identified as the the critical
rigidity, $\mu_c$. See also Fig. 2.
} 
\\\\
{\it Low-lying eigenvalues near criticality:} In general, there will exist a
critical value of the rigidity, $\mu_n^\pm$,  for each mode  
at which  the eigenvalue
$\lambda_{\mu,n}^\pm$ turns positive and above which it remains
so.  Whereas short wavelength modes stay positive once they turn positive,  as we have just seen, their misbehaved long wavelength counterparts need not remain positive.  The physically significant rigidity, however, is the value where the eigenvalue turns positive for the last time.  We do find that these values are ordered:
$\cdots <\mu_2^+< \mu_2^-<\mu_1^+ <\mu_1^-$;  
the last eigenvalue to turn and stay positive is $\lambda_{\mu,1}^-$, 
and we identify $\mu_1^-$ as $\mu_c$, the rigidity controlling the stability of the arch. 
\\\\
(i) Above $\mu_c$,  
the number of nodes correlates with energy:
$0<\lambda_{\mu,1}^-<
\lambda_{\mu,1}^+ <\lambda_{\mu,2}^- < \lambda_{\mu,2}^+ <
\cdots$; the odd state with a single node
$|\lambda_{\mu,1}^-\rangle$ is the stable ground state. 
There are no eigenvalue crossovers once the arch turns stable.
\\
(ii)
Within the sub-critical region, $\mu<\mu_c$, multiple eigenvalue  crossovers will  generally occur between long wavelength  modes. However, 
for each pair of eigenvalues, there will be a final crossover.
The last crossover to occur is between 
$\lambda_1^-$ and $\lambda_1^+$, occurring when $\mu=\mu_{1}^{+-}$.  The dependence of these two eigenvalues 
on $\mu$ is represented in Fig. 5 for the aspect ratio $H/L=1/5$.  
\\\\
We observe the following behavior in Fig. 5:
\\\\
(i) If $\mu>\mu_{1}^{+-}$,  then $\lambda_1^-<\lambda_1^+$,  and   
$|\lambda_{\mu,1}^-\rangle$ is the ground state, albeit unstable below $\mu_c$.  
If, in addition, $\mu> \mu_1^+$ (where $\lambda_1^+$ turns positive), 
then $|\lambda_{\mu,1}^-\rangle$  is also the unique mode of instability.
If $\mu$ falls below $\mu_{1}^+$, however, $\lambda_{\mu,1}^+$ turns negative 
so that $|\lambda_{\mu,1}^+\rangle$ provides an additional unstable mode. The dominant mode remains $|\lambda_{\mu,1}^-\rangle$.\\\\
(ii) As $\mu$ is lowered below the crossover value $\mu_1^{+-}$,  however,
then $\lambda_{\mu,1}^+ < \lambda_{\mu,1}^-$ so that
$|\lambda_{\mu,1}^+\rangle$
replaces $|\lambda_{\mu,1}^-\rangle$ as the ground state. The parity of the 
dominant unstable mode changes, with an additional node. 
The changing parity of the dominant instability was not anticipated.
 \\\\
{\it Physical representation of the near-critical unstable mode:}
 If $\mu_1^+ < \mu<\mu_c$, we saw that the single unstable mode is 
$|\lambda_1^-\rangle$. This represents a
symmetric buckling of a catenary,
with two  {\it nodes}  as illustrated in Fig. 6.
\begin{center}
\includegraphics[scale=0.45]{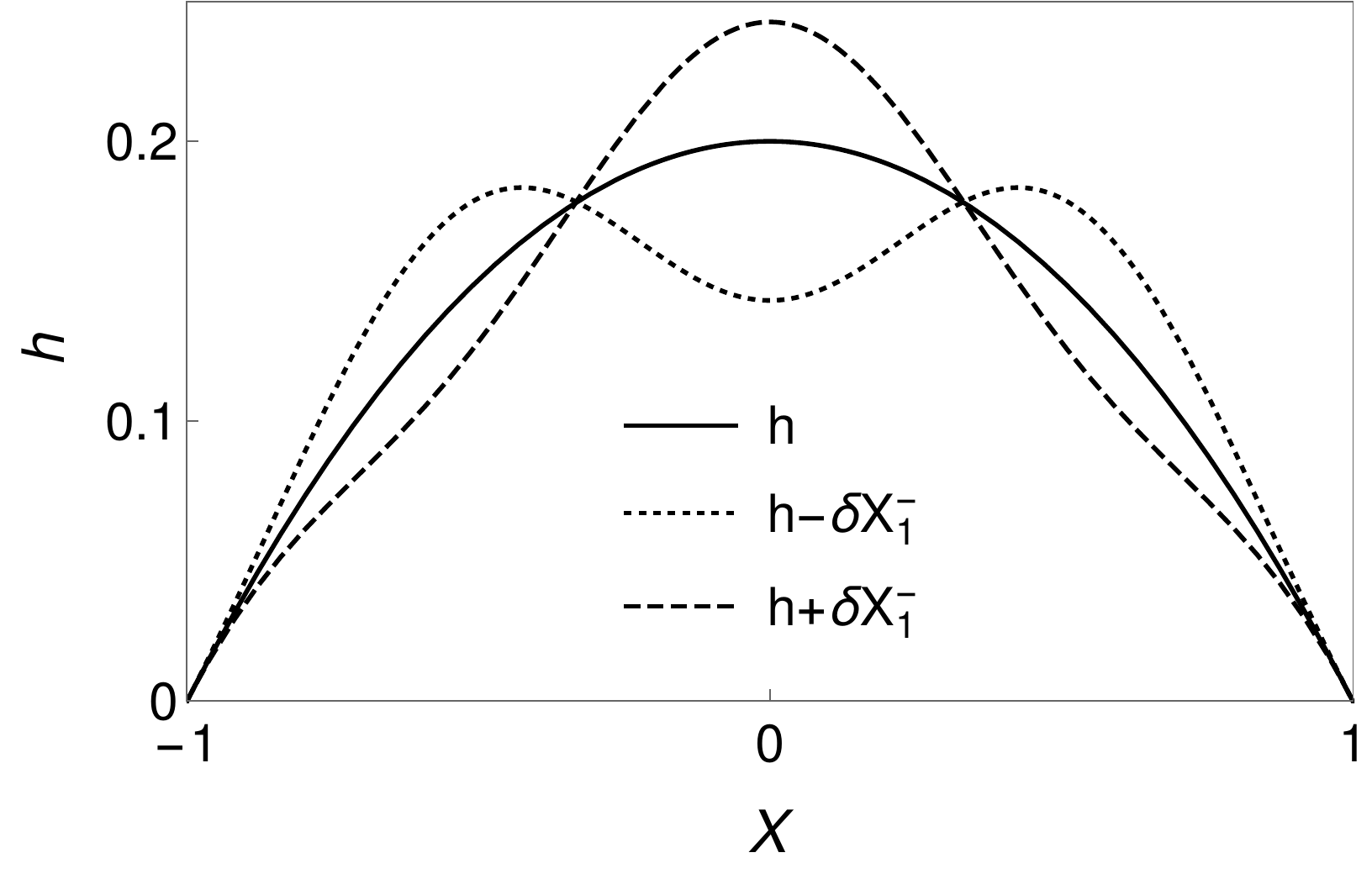}
\end{center}
{\small {\bf FIG. 6:} Near-critical physically distinct degenerate unstable modes represented by the ground state $|\lambda_{\mu,1}^-\rangle$: 
$\mathbf{X}\pm \delta \mathbf{X}$.
The length of the deformed curve has been increased for clarity.}
\\\\ 
The use of italics requires explanation. We know that 
the single tangential node in the ground state implies that the 
corresponding geometrical deformation has two nodes in the normal deformation.
This is a straightforward consequence of the lineared isometry
constraint, $\Psi'+ \kappa \Phi=0$. 
But this equation also implies that the nodes of $\Psi$ never coincide with those of $\Phi$. As a result, the catenary 
possesses no fixed points ($\delta {\bf X}\ne 0$) except at its
boundaries. Note also that an initially collapsing arch must necessarily rise somewhere.  It cannot fall all together at once,  a consequence of the isometry constraint and the boundary conditions. 

As $\mu$ is lowered further, 
and bending energy becomes insufficient to arrest its collapse
on increasingly shorter wavelengths, the perturbed catenary
will initially rise and fall  an increasing number of times
number of nodes, reflecting the nodal behavior of the most unstable mode, as well as  the increased number of modes of instability.

Physically, each unstable mode describes two distinct geometrical modes of collapse: the spectral analysis does not distinguish energetically
between the positive and negative deformations illustrated in Fig.
6. But buckling up is not the same as buckling down. To resolve
this quadratic degeneracy, one needs to
expand the energy out to third order in deformations, a
calculation complicated by the necessity to treat the isometry
constraint quadratically. Experiments with
thin sheets confirm that the lower energy state is alway the one
falling at its center.
 
There are few extant higher order calculations, even without
local constraints to contend with. One example we are aware  of 
involves global rather than local constraints \cite{HelfOuYang}. And, to the best of our knowledge, an independent confirmation of this result has yet to be performed.

This perturbative analysis does not discount the possibility of
the subcritical arch relaxing into a new equilibrium stabilized
by bending energy. Unless such a state exists, tension will
build up at the antinodes as the deformation grows. This
would lead to the disintegation of a masonry arch at these {\it
hinges} and the subsequent free fall of the fragments.

\section{Conclusions and future work}
In this paper, we have introduced a geometrical framework to assess the stability of a free-standing thin-walled structure. This involves
understanding how surfaces deform isometrically, the
identification of the appropriate bending energy and the
spectral analysis of the self-adjoint 
operator controlling perturbations about equilibrium. 
While geometry was well-known to play a role in our identification of equilbrium its role in understanding the behavior of deformations about equilibrium does not appear to have been appreciated, never mind explored in this limit.  In general there is, of course, a lot more to solid mechanics  than is captured by our phenomenolical model \cite{Steigmann}.  One should think of this energy as a phenomenological one, where the emphasis is on capturing the essential features of the problem.  Whille deriving the appropriate bending energy from first principles is likely to be no simple matter, when the dust settles one would  expect it to look not unlike the energy written down in Eq.(\ref{HB}), with 
spontaneous curvatures determined---consistent with the protocol--by the 
equilibrium reference values of the curvature.  If, however, the thin wall approximation does not
apply, our geometrical approach is rendered invalid and it is necessary to approach the subject very differently.


We have illustrated this framework using the historically significant example of a catenary arch.
The example itself may appear quaint; its analysis does, however, illustrate the extraordinarily sensitive dependence of the stability on the rigidity.  

It should be stressed that we have not attempted to capture every feature of the linear stability of a catenary arch much less attempt to examine any more complicated geometry. Understanding the complex erratic behavior of the spectrum of the self-adjoint operator controlling the response, alone, would keep us occupied for a while.  
We have, however, described a number of the essential elements of the problem,  justiifying not just why the Hookean construction works---something  every mason knows--- but also providing a framework to predict if an arch will stand and how it fails if it does not.
We hope our work will inspire others to pursue these issues.

The natural next step is to examine the stability of genuine
three-dimensional structures. Indeed, a complete treatment of the catenary vault should rightly
accommodate deformations that break the cylindrical symmetry,
deforming the arch into a more general metrically flat tangent developable surface \cite{doCarmo} consistent with the boundary conditions. 
As it turns out, 
when the rigidity is isotropic as we have supposed for simplicity, such modes of failure never form the ground state so treating them does not change our conclusions. If, however, the wall is constructed using an anisotropic material, so that $H_B$ in Eq.(\ref{HB}) is replaced by something like
\begin{equation}H_B = \frac{1}{2}\, \sum_{\small I}\mu_{\small I}\, \int dA\, (C_I-{\cal C}_I(u))^2 \,,
\label{HB1}
\end{equation} 
with different moduli along orthogonal directions,
the dominant mode of instability would be expected to lie along the transverse direction if bending is soft along this direction, or equivalently, $\mu_2/\mu_1$ is sufficiently small.

In general, the stability of a three-dimensional structure will depend very sensitively both on its Gaussian curvature as well as the boundary conditions.
A mathematical surface with positive Gaussian curvature, 
such as a closed dome on a fixed base, does not support any small isometric deformations \cite{Spivak}. So
if stretching is not accommodated, one would expect isometry alone  to stabilize the structure. This may be true of pingpong balls. The fault with
this reasoning here is that vertical cracks will develop towards the base of a masonry dome where the
structure becomes subject to substantial tension (or hoop stresses) along the parallels. The two sides of the crack
will behave as free boundaries that become free to move independently under deformation. 
The stability of the dome needs to be reassessed accommodating the free boundaries along these cracks. 
It is a sobering thought that 
a reliable geometrical framework to accomodate free boundaries on thin walls remains to be developed. But that is another story.
\\\\
{\it Acknowledgments:}
We benefitted from discussions with L Mahadevan during the early stages of this work. Indeed, it was he who pointed out that the stability of masonry was a question  that was due reappraisal in the thin wall limit. We also thank James Hanna, Martin M\"uller and David Steigmann for useful comments. This work was supported in part by CONACyT grant no. 180901 to JG.


\begin{thebibliography} {99}
\bibitem{Roman}
 In a semi-circular Roman arch, the wayward thrust is tamed by the  weight distributed above it.
\bibitem{Displacements}   Our focus is on the role of bending energy in arresting the collapse of an otherwise gravitationally unstable configuration. We do not examine its stability under the displacement of its support.
\bibitem{basis}
We choose as our basis (expressed as functions of $X$ rather than $s$)
the  funtions which result from the Gram-Schmidt  orthogonalization of the following sums of Fourier modes:
{\small 
\[
f_n^+(X)=\frac{1}{\sqrt{L}}\left(\cos\frac{(2n+1)\pi X}{2L}+\frac{2n+1}{2n+3}\cos\frac{(2n+3)\pi X}{2L}\right)\,,
\]
}
and
{\small
\[f_n^-(X)=\frac{1}{\sqrt{L}}\left(\sin \frac{n\pi x}{L}+\frac{n}{n+1}\sin \frac{(n+1)\pi X}{L}\right)\,.
\]
}
These modes are consistent with the boundary conditions: $f_n,f_n',f_n''=0$ on $X=\pm L$. 
\\\\
\end{thebibliography}
\end{document}